\documentclass[Physsubmission, Phys]{SciPost}

\hypersetup{
    colorlinks,
    linkcolor={red!50!black},
    citecolor={blue!50!black},
    urlcolor={blue!80!black}
}

\usepackage[bitstream-charter]{mathdesign}
\DeclareSymbolFont{usualmathcal}{OMS}{cmsy}{m}{n}
\DeclareSymbolFontAlphabet{\mathcal}{usualmathcal}

\begin{document}
\begin{center}{\Large \textbf{
Multiscale pentagon integrals to all orders\\
}}\end{center}

\begin{center}
Dhimiter D. Canko\textsuperscript{1,2},
Costas G. Papadopoulos\textsuperscript{1} and
Nikolaos Syrrakos\textsuperscript{1,3,4$\star$}
\end{center}

\begin{center}
{\bf 1} Institute of Nuclear and Particle Physics, NCSR `Demokritos', Agia Paraskevi, 15310, Greece
\\
{\bf 2} Department of Physics, University of Athens,\\Zographou 15784, Greece
\\
{\bf 3} Physics Division, National Technical University of Athens, \\Athens 15780, Greece
\\
{\bf 4} Physik-Department, Technische Universität München,\\ James-Franck-Str. 1, 85748 Garching, Germany\\
*syrrakos@inp.demokritos.gr
\end{center}

\begin{center}
\today
\end{center}


\definecolor{palegray}{gray}{0.95}
\begin{center}
\colorbox{palegray}{
  \begin{tabular}{rr}
  \begin{minipage}{0.1\textwidth}
    \includegraphics[width=35mm]{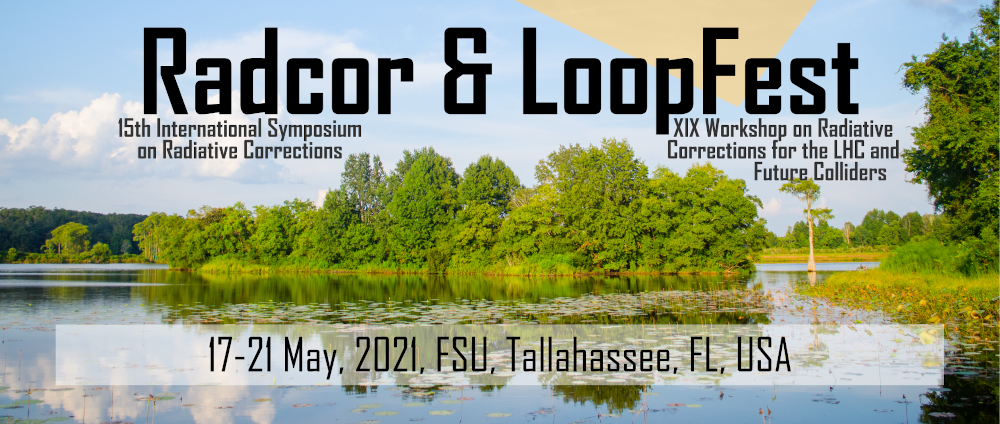}
  \end{minipage}
  &
  \begin{minipage}{0.85\textwidth}
    \begin{center}
    {\it 15th International Symposium on Radiative Corrections: \\Applications of Quantum Field Theory to Phenomenology,}\\
    {\it FSU, Tallahasse, FL, USA, 17-21 May 2021} \\
    \doi{10.21468/SciPostPhysProc.?}\\
    \end{center}
  \end{minipage}
\end{tabular}
}
\end{center}

\section*{Abstract}
{\bf
We present analytical results for one-loop five-point master integrals with up to three off-shell legs. The method of canonical differential equations along with the Simplified Differential Equations approach is employed. All necessary boundary terms are given in closed form, resulting to solutions in terms of Goncharov Polylogarithms of arbitrary weight. Explicit results up to weight six will be presented.
}

\vspace{10pt}
\noindent\rule{\textwidth}{1pt}
\tableofcontents\thispagestyle{fancy}
\noindent\rule{\textwidth}{1pt}
\vspace{10pt}

\section{Introduction}
\label{sec:intro}
Our best understanding of Nature in its most fundamental level is encoded in the Standard Model (SM) of particle physics, written in the mathematical language of Quantum Field Theory. Currently, experimental data coming mostly from the realm of Cosmology, such as the existence of dark matter, impose major challenges on our fundamental theories, since they are not predicted by the SM. Thus, we have concrete experimental signs that the SM does not suffice to explain Nature at its most fundamental level.

On the other hand, the SM reigns supreme when we compare its predictions to experimental data coming from the LHC. Since collider physics remains the best way to explore the validity of the SM predictions against experimental data, a new precision program has been initiated in order to test our current understanding of particle physics with the highest possible precision, both from an experimental and a theoretical point of view \cite{Heinrich:2020ybq}. Our hope is that through the careful comparison of highly precise measured cross sections against equally precise theoretical predictions, deviations from the SM will be discovered in collider experiments, leading to constraints on its possible extensions \cite{Amoroso:2020lgh}. 

It is estimated that the Run 3 of the LHC as well as its expected High Luminosity upgrade will require from a theoretical standpoint at least Next-to-Next-to-Leading Order (NNLO) corrections to the QCD dominated processes \cite{Amoroso:2020lgh}. The current precision frontier at NNLO lies at $2\to3$ processes. A key factor in the determination of theoretical predictions for these processes is the calculation of the relevant two-loop Feynman integrals (FI). Despite the ever increasing sophistication of computational packages such as \texttt{FIESTA4} \cite{Smirnov:2015mct} and \texttt{pySecDec} \cite{Borowka:2017idc} dedicated to the numerical evaluation of FI, analytic results for FI are still important, especially for physical regions of the phase space, where in many cases a direct numerical evaluation is not possible.

Working in the framework of dimensional regularisation in $d=4-2\epsilon$ space-time dimensions, it can be shown that FI satisfy so-called Integration-By-Parts (IBP) identities \cite{Chetyrkin:1981qh}. These relations allow us to obtain a minimal set of FI known as Master integrals (MI), that need to be computed for a specific scattering process. Regarding MI relevant for $2\to3$ NNLO calculations, all MI involving massless particles are known \cite{Papadopoulos:2015jft,Gehrmann:2015bfy,Chicherin:2017dob,Gehrmann:2018yef,Chicherin:2018mue,Abreu:2018rcw,Chicherin:2018old} and implemented in the \texttt{C++} library \texttt{pentagon functions} \cite{Chicherin:2020oor}, all planar MI for processes involving one off-shell leg have recently been calculated using a numerical \cite{Abreu:2020jxa} and an analytical \cite{Canko:2020ylt} approach and even more recently there has been important progress for some of the non-planar topologies involving one off-shell leg, known as \textit{hexaboxes} \cite{Papadopoulos:2019iam,Abreu:2021smk}.

Despite these very important accomplishments, at some point we will have to consider processes involving more external massive legs and massive propagators. This will require the calculation of very complicated two-loop MI. In order to gauge the level of mathematical complexity that these MI will present, it is instructive to study first their one-loop counterparts. To that end, in this contribution we will present analytical results for one-loop five-point MI with up to three off-shell legs and massless internal lines. All results are given in terms of Goncharov Polylogarithms (GPLs), a class of special functions which is well understood by now \cite{Goncharov:1998kja,Duhr:2011zq,Duhr:2012fh,Duhr:2014woa}, up to transcendental weight six, although the computational approach that was used allows one to trivially obtain higher weight analytical results.

\section{Computational framework}
The modern approach for computing MI analytically is through the use of the method of differential equations (DE) \cite{Henn:2014qga,de1,de2,de3,de4}. After using IBP identities and identifying a basis $\textbf{G}$ of MI, one differentiates this basis with respect to all kinematic invariants
\begin{equation}
       \frac{\partial}{\partial s_{ij}} \textbf{G} = \textbf{A}(\{s_{ij}\},\epsilon) \textbf{G}
\end{equation}
In general the matrix $\textbf{A}$ can be very complicated. The introduction of the canonical DE \cite{Henn:2013pwa} brought forth a revolution in the computation of MI \cite{Kotikov:2021tai}. This new approach suggests that instead of using basis $\textbf{G}$, one can use a special basis, known as a pure basis of MI, $\textbf{g} = \textbf{T} \textbf{G}$ for which the DE has the following form, known as \textit{canonical form},
\begin{equation}
\mathrm{d} \textbf{g} = \epsilon \sum\limits_a {\mathrm{d}\log \left( {{W_a}} \right){\tilde{\textbf{M}}_a} \textbf{g}} 
    \label{eq:canonical}
\end{equation}
The functions $W_a$ are known as \textit{letters} of the so-called \textit{alphabet}, which is the set of all $W_a$ for a specific family of MI. When $W_a$ are \textit{rational} functions of the differential variables, \eqref{eq:canonical} is solved by recursively integrating order-by-order in $\epsilon$ in terms of GPLs, that can be defined as iterated integrals in the following way
\begin{align}
        \mathcal{G}(a_1,a_2,\ldots, a_n;x) &= \int_0^x \, \frac{\mathrm{dt}}{t-a_1}\mathcal{G}(a_2,\ldots, a_n;t)\\
        \mathcal{G}(0,\ldots, 0;x) &= \frac{1}{n!}\log^n(x)
\end{align}

For five-point MI, achieving an analytical solution in terms of GPLs beyond weight three is a non-trivial task. This is due to the fact that several of the letters $W_a$ are algebraic functions of the differential variables, thus prohibiting a direct integration in terms of GPLs. More specifically, these algebraic letters consist of square roots arising from leading singularities of massive three-point functions and the Gram determinant of the five-point external momenta. 

A variant of the standard DE method, known as Simplified Differential Equations (SDE) approach \cite{Papadopoulos:2014lla}, has been shown to effectively circumvent the problem of algebraic letters in many cases \cite{Canko:2020ylt,Syrrakos:2020kba,Syrrakos:2021nij}, thus allowing fully analytical solutions in terms of GPLs to be achieved. In the SDE approach, we introduce an external parameter $x$ in the external momenta and derive the DE by differentiating with respect to only that parameter, regardless of the number of scales of the scattering process. When this approach is applied on a pure basis $\textbf{g}$ of MI, a canonical SDE can be derived. In many cases the new letters $W'_b$ are fully rationalised in $x$,  $W'_b = x-l_b$, yielding the form
\begin{equation}
\partial_x \textbf{g} = \epsilon \sum\limits_b {\frac{1}{{x - {l_b}}}{\textbf{M}_b}\, \textbf{g}} 
    \label{eq:canonicalx}
\end{equation}
In what follows we will refer to $l_b$ as letters in the SDE approach. All kinematic dependence is included in these $l_b$ functions, living the residue matrices $\textbf{M}_b$ to consist solely of rational numbers. The form of \eqref{eq:canonicalx} allows for its solution to be trivially expressed in terms of GPLs, assuming the necessary boundary terms are obtained. To do so, we employ the method of expansion-by-regions \cite{Jantzen:2012mw}, with which we compute the $x\to0$ limit for each pure basis element. 

\section{Results}

\subsection{Integral families}
\begin{figure}[h!]
    \centering
    \includegraphics[width=4cm]{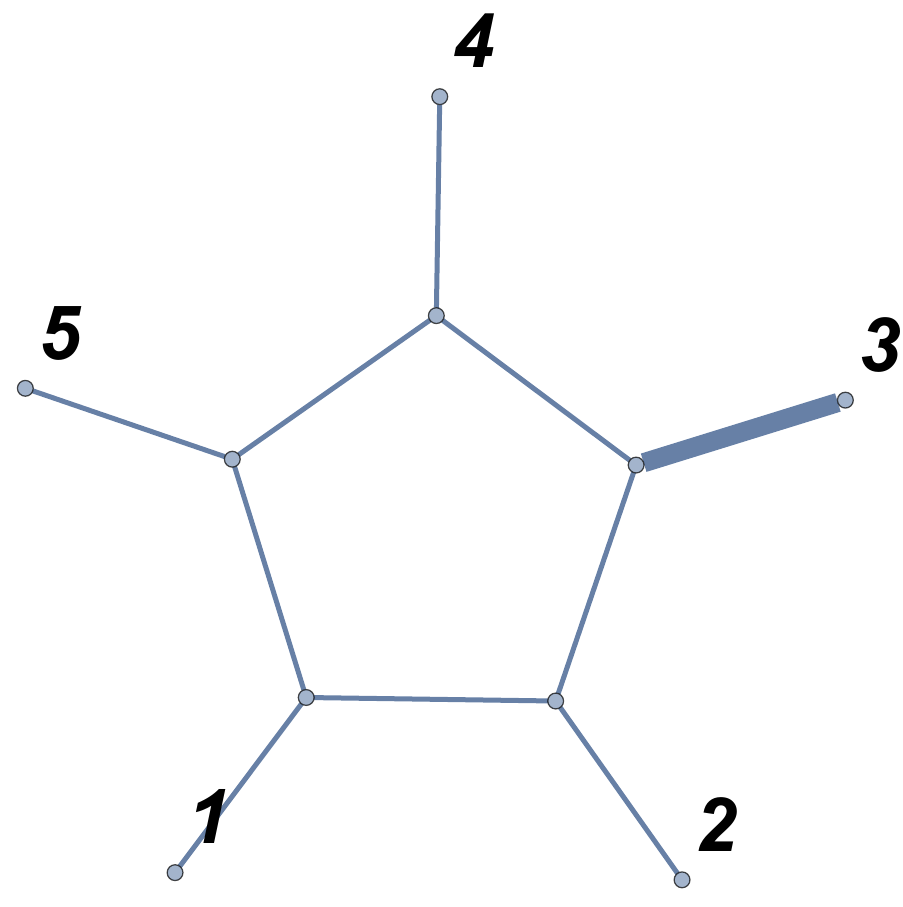}
    \includegraphics[width=4cm]{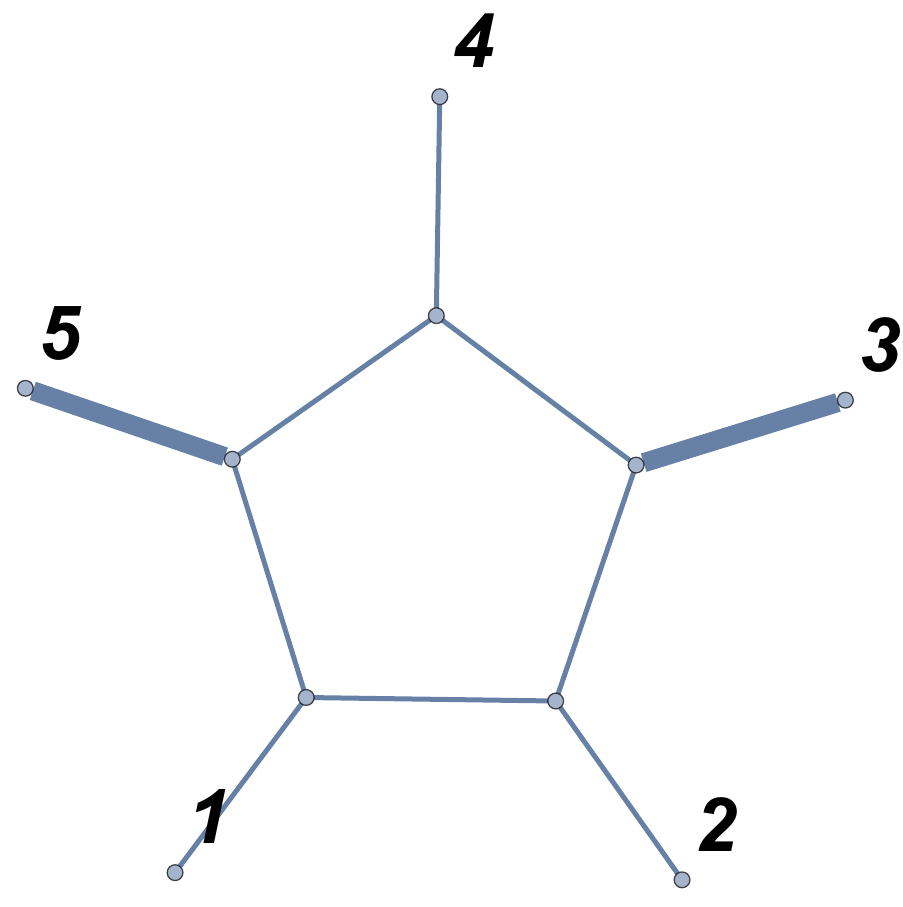}
    \includegraphics[width=4cm]{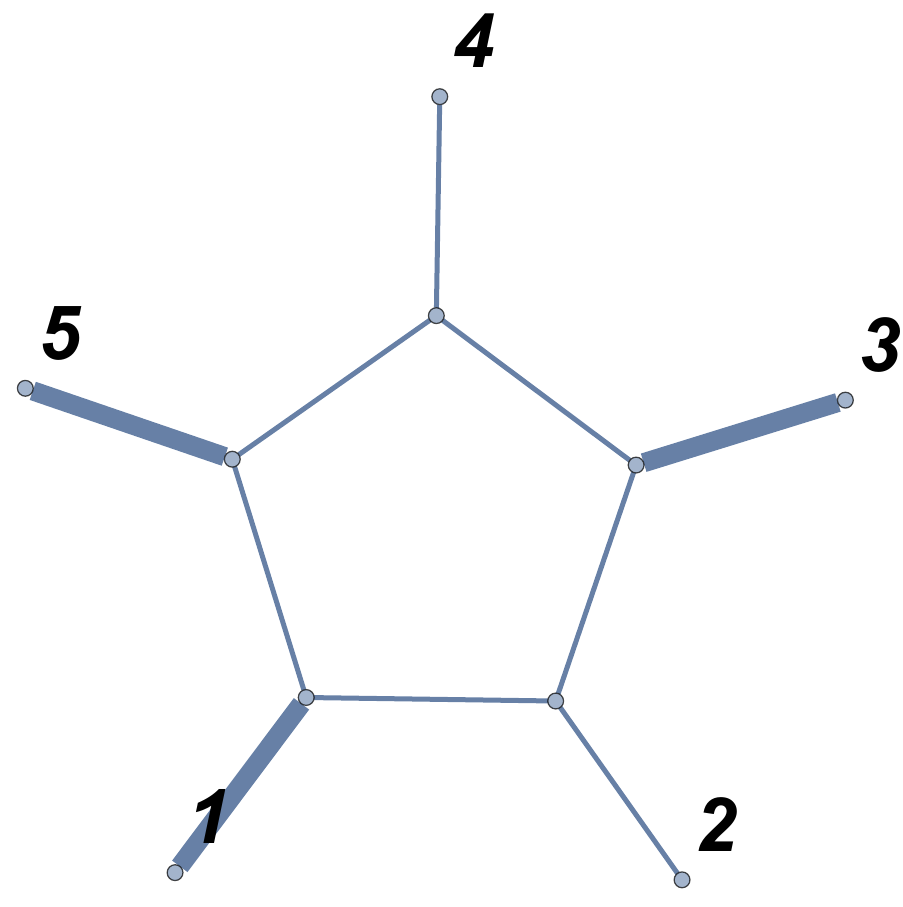}
    \caption{Top-sector MI for the pentagon families considered in this contribution.}
    \label{fig:graphs}
\end{figure}

The integral families considered in this contribution, whose top-sector diagrams are depicted in figure \ref{fig:graphs}, are defined as follows
\begin{equation}\label{eq:pentaf}
    G_{a_1a_2a_3a_4a_5} = \int~ \frac{\mathrm{d}^d k_1}{i\pi^{(d/2)}}~ \frac{\mathrm{e}^{\epsilon \gamma_{E}}}{\mathcal{D}_1^{a_1}\mathcal{D}_2^{a_2}\mathcal{D}_3^{a_3}\mathcal{D}_4^{a_4}\mathcal{D}_5^{a_5}},\quad d=4-2\epsilon
\end{equation}
with
\begin{align}
    &\mathcal{D}_1=-(k_1)^2,~ \mathcal{D}_2=-(k_1+q_1)^2,~ \mathcal{D}_3=-(k_1+q_1+q_2)^2 \nonumber \\
    &\mathcal{D}_4=-(k_1+q_1+q_2+q_3)^2,~ \mathcal{D}_5=-(k_1+q_1+q_2+q_3+q_4)^2
\end{align}
The kinematics for the three families are as follows,
\begin{itemize}
    \item 1-mass: $\sum_{i=1}^5q_i=0$, $q_i^2=0$, $i=1,2,4,5$, $q_3^2=m_3^2$
    \item 2-mass: $\sum_{i=1}^5q_i=0$, $q_i^2=0$, $i=1,2,4$, $q_3^2=m_3^2,q_5^2=m_5^2$
    \item 3-mass: $\sum_{i=1}^5q_i=0$, $q_i^2=0$, $i=2,4$, $q_1^2=\bar{m}_1^2,q_3^2=m_3^2,q_5^2=m_5^2$
\end{itemize}
We introduce the following $x$-parametrization\footnote{We use the abbreviations $p_{ij}=p_i+p_j$ and $p_{ijk}=p_i+p_j+p_k$ and similarly for $q$ later.}
\begin{equation}\label{eq:xparm}
    q_1=xp_1,~q_2=xp_2,~q_3=p_{123}-xp_{12},~q_{4}=p_{4},~q_5=-p_{1234}
\end{equation}
The kinematics in this underline momentum parametrization is
\begin{itemize}
    \item 1-mass: $\sum_{i=1}^5p_i=0$, $p_i^2=0$, $i=1,2,3,4,5$
    \item 2-mass: $\sum_{i=1}^5p_i=0$, $p_i^2=0$, $i=1,2,3,4$, $p_5^2=m_5^2$
    \item 3-mass: $\sum_{i=1}^5p_i=0$, $p_i^2=0$, $i=2,3,4$, $p_1^2=m_1^2,p_5^2=m_5^2$
\end{itemize}
Introducing \eqref{eq:xparm} results in a mapping between the kinematic invariants in the original momentum parametrization, $q_i$, and the underline momentum parametrization $\{x,p_i\}$ for each of the three pentagon families.\footnote{We use the abbreviations $s_{ij}=q_{ij}^2,~ S_{ij}=p_{ij}^2$.}
\begin{align}\label{eq:mommap}
    \text{1-mass}:&~ s_{12}= S_{12} x^2,s_{23}= S_{23} x-S_{45} x+S_{45},s_{34}= x \left(S_{12}
   (x-1)+S_{34}\right),s_{45}= S_{45},\nonumber \\
   &s_{15}= S_{15} x,m_3^2= (x-1) \left(S_{12}
   x-S_{45}\right)\\
    \text{2-mass}:&~ m_3^2= (x-1) \left(S_{12} x-S_{45}\right),s_{12}= S_{12} x^2,s_{23}= S_{23}
   x-S_{45} x+S_{45}\nonumber \\
   &s_{34}= m_5^2 (-x)+m_5^2+x \left(S_{12}(x-1)+S_{34}\right),s_{45}=S_{45},s_{15}= m_5^2 (-x)+m_5^2+S_{15} x \nonumber \\
   \text{3-mass}:&~ s_{12}= S_{12} x^2,s_{23}= x \left(m_1^2 (x-1)+S_{23}\right)-S_{45}
   x+S_{45},\nonumber \\
   &s_{34}= m_5^2 (-x)+m_5^2+x \left(S_{12} (x-1)+S_{34}\right),
   s_{45}=
   S_{45},\nonumber\\
   &s_{15}= x \left(m_1^2 (x-1)+S_{15}\right)+m_5^2 (-x)+m_5^2,\bar{m}_1^2= m_1^2
   x^2,m_3^2= (x-1) \left(S_{12} x-S_{45}\right) \nonumber 
\end{align}

\subsection{Differential equations}
Constructing pure bases for the pentagon families under consideration is by now a trivial exercise. Following the ideas of \cite{Abreu:2018rcw, Abreu:2020jxa}, the top sector basis element at the \textit{integrand} level is of the form
\begin{equation}\label{eq:uttop}
    \epsilon^2 \frac{\mathcal{P}_{11111}}{\sqrt{\Delta_5}} \Tilde{G}_{11111}
\end{equation}
where $\mathcal{P}_{11111}$ is the Baikov polynomial corresponding to the top sector \textit{integral} $G_{11111}$ for each family, $\Tilde{G}_{11111}$ is the top sector \textit{integrand} of each family  and $\Delta_5 = \det[q_i\cdot q_j]$ is the Gram determinant of the five-point external momenta. The remaining pure basis elements can be constructed through the study of the leading singularities of their corresponding diagrams \cite{Henn:2014qga}. Using Azurite \cite{Georgoudis:2016wff} and Kira2 \cite{Klappert:2020nbg} we can identify 13, 15 and 18 MI for the 1-mass, 2-mass and 3-mass pentagon families respectively.

When considering five-point scattering with massless propagators, a number of square roots of the kinematic invariants enter the DE of the corresponding pure bases. These square roots originate from leading singularities of massive three-point functions, which are represented by square roots of the K\"allen function $\lambda(x,y,z) = x^2-2 x y-2 x z+y^2-2 y z+z^2$ and from square roots of the Gram determinants of the five-point external momenta. The existence of these square roots makes the task of solving these canonical DE in terms of GPLs quite challenging, and in some cases even impossible \cite{Brown:2020rda}.

It turns out however, that for the families considered in this contribution, the SDE approach can overcome these difficulties. Introducing the $x$-parametrization \eqref{eq:xparm} rationalises all square roots in terms of the differential variable $x$, allowing us to integrate the canonical DE and express the final result in terms of GPLs. The canonical SDE for each of the three pentagon families has the following form,
\begin{equation}\label{eq:cande}
\partial_{x} \textbf{g}=\epsilon \left( \sum_{i=1}^{l_{max}} \frac{\textbf{M}_i}{x-l_i} \right) \textbf{g}
\end{equation}
where $\textbf{g}$ is the pure basis for each family, $\textbf{M}_i$ are the residue matrices corresponding to each letter $l_i$ and $l_{max}$ is the length of the alphabet in $x$. The length of the alphabet for each of the three families considered in this subsection is $l_{max}^{1m}=11,~l_{max}^{2m}=14,~l_{max}^{3m}=19$. For more information on the structure of these alphabets, we refer the interested reader to \cite{Syrrakos:2020kba,Syrrakos:2021nij}.

The last ingredient that is missing to solve \eqref{eq:cande} are the boundary terms. To find them we follow the techniques developed in \cite{Canko:2020gqp,Canko:2020ylt}. We define the \textit{resummation matrix} $\textbf{R}$ as follows
\begin{equation}\label{eq:resm0}
    \textbf{R} =\textbf{S} e^{\epsilon \textbf{D} \log(x)} \textbf{S}^{-1}
\end{equation}
where $\textbf{S}$ and $\textbf{D}$ are matrices coming from the Jordan decomposition of the residue matrix corresponding to $l_1=0$, $\textbf{M}_1=\textbf{S} \textbf{D} \textbf{S}^{-1}$. Having expressed the pure bases in terms of MI, $\textbf{g} = \textbf{T} \textbf{G}$, we use the expansion-by-regions method implemented in the \texttt{asy} code which is shipped along with \texttt{FIESTA4}~\cite{Smirnov:2015mct}, to find their asymptotic limit for $x\to0$.
\begin{equation}\label{eq:regions}
    {G_i}\mathop  = \limits_{x \to 0} \sum\limits_j x^{b_j + a_j \epsilon }G^{(b_j + a_j \epsilon)}_{i} 
\end{equation}
where $a_j$ and $b_j$ are integers and $G_i$ are the individual members of the basis $\textbf{G}$. As explained in \cite{Canko:2020ylt}, we can construct the relation
\begin{equation}\label{eq:bounds}
    \mathbf{R} \mathbf{b}=\left.\lim _{x \rightarrow 0} \mathbf{T} \mathbf{G}\right|_{\mathcal{O}\left(x^{0+a_{j} \epsilon}\right)}
\end{equation}
where $\textbf{b}=\sum_{i=0}^n\epsilon^i\textbf{b}_0^{(i)}$ are the boundary terms that we need to compute. The right-hand-side of  \eqref{eq:bounds} implies that, apart from the terms $x^{a_i  \epsilon}$ coming from \eqref{eq:regions}, we expand around $x=0$, keeping only terms of order $x^0$. This procedure allows us to fix all the necessary boundary terms in closed form, thus allowing us to obtain analytical solutions of \eqref{eq:cande} in terms of GPLs of arbitrary weight.

\subsection{Solutions}
In this contribution we provide solutions up to weight six for each of the considered pentagon families, which can be written in the following compact form,
\begin{equation}
\label{eq:solution}
\begin{split}
\textbf{g}&=\epsilon^0 \textbf{b}_0^{(0)}+\epsilon \left(\sum {\cal G}_i \textbf{M}_i \textbf{b}_0^{(0)}+\textbf{b}_0^{(1)}\right)+\epsilon^2 \left(\sum {\cal G}_{ij} \textbf{M}_i\textbf{M}_j\textbf{b}_0^{(0)}+\sum {\cal G}_i \textbf{M}_i \textbf{b}_0^{(1)}+\textbf{b}_0^{(2)} \right)+ \dots \\  
&+ \epsilon^6 \left(\textbf{b}_0^{(6)}+ \sum {\cal G}_{ijklmn} \textbf{M}_i \textbf{M}_j \textbf{M}_k \textbf{M}_l \textbf{M}_m \textbf{M}_n \textbf{b}_0^{(0)} + \sum {\cal G}_{ijklm} \textbf{M}_i \textbf{M}_j \textbf{M}_k \textbf{M}_l \textbf{M}_m \textbf{b}_0^{(1)} \right. \\
&+\left. \sum {\cal G}_{ijkl} \textbf{M}_i \textbf{M}_j \textbf{M}_k \textbf{M}_l \textbf{b}_0^{(2)} +\sum {\cal G}_{ijk} \textbf{M}_i\textbf{M}_j\text{M}_k \textbf{b}_0^{(3)}+\sum {\cal G}_{ij} \textbf{M}_i\textbf{M}_j\textbf{b}_0^{(4)}+\sum {\cal G}_i \textbf{M}_i \textbf{b}_0^{(5)} \right) 
\end{split}
\end{equation}
were $\mathcal{G}_{ab\ldots}:= \mathcal{G}(l_a,l_b,\ldots;x)$ represent the GPLs. The $b_0^{(i)}$ terms, with $i$ indicating the corresponding weight, consist of Zeta functions $\zeta(i)$, logarithms and GPLs of weight $i$ which have as arguments rational functions of the underline kinematic variables $S_{ij}$.

For all pentagon families we have made heavy use of the \texttt{Mathematica} package \texttt{PolyLogTools} \cite{Duhr:2019tlz} for the manipulation of the resulting GPLs. In Tables \ref{tab:polylog1} and \ref{tab:polylog2} we provide an analysis of our results for each family, regarding the number of GPLs that appear in each transcendental weight, where the weight is counted as the number of $l_i$ indices of $\mathcal{G}(l_a,l_b,\ldots;x)$. These numbers are obtained by gathering all GPLs that appear up to order $\mathcal{O}(\epsilon^6)$ in each integral family, and distinguishing them according to their corresponding weight. For comparison, we perform the same task for the top-sector basis element of each family.

\begin{table}[ht]
\centering
\begin{tabular}{|c|c|c|c|c|}
\hline
 Family     & W=1 & W=2 & W=3 & W=4 
 \\
 \hline 
$P_{1m}$      & 10 (2)   & 50 (21) & 170 (99) & 496 (339) 
\\
\hline
$P_{2m}$      & 9 (0)   & 54 (16) & 204 (106) & 628 (406)  
\\
\hline
$P_{3m}$     & 13 (0)  & 87 (24) & 349 (172) & 1115 (696) 
\\
\hline
\end{tabular}
\caption{Number of GPLs entering in the solution. Top-sector b.e. in parenthesis.}
\label{tab:polylog1}
\end{table}

\begin{table}[ht]
    \centering
\begin{tabular}{|c|c|c|}
\hline
 Family     & W=5 & W=6  \\
 \hline 
$P_{1m}$      & 1322 (959)   & 2983 (1924) \\
\hline
$P_{2m}$      & 1728 (1254)   & 4341 (2656)  \\
\hline
$P_{3m}$     & 3145 (2228)  & 7849 (4656)  \\
\hline
\end{tabular}
\caption{Number of GPLs entering in the solution. Top-sector b.e. in parenthesis.}
\label{tab:polylog2}
\end{table}

We observe a huge increase in the number of GPLs beyond weight 5, in comparison to lower weights. Despite the fact that the top sector basis elements appear to have lower-weight GPLs, we note that each top-sector basis element starts from $\mathcal{O}(\epsilon^3)$. We also provide numerical results and timing obtained using \texttt{handyG} \cite{Naterop:2019xaf,Vollinga:2004sn} for the top-sector basis element of each family for a Euclidean point in table \ref{tab:polylog3}.

\begin{table}[ht]
\centering
\begin{tabular}{|c|c|c|}
\hline
 Top-Sec     & Time (sec) & Result  \\
 \hline 
$g_{13}$      & 1.90759   & $0.0944261 \epsilon^3 + 0.31615 \epsilon^4 + 0.666923 \epsilon^5 + 1.09948 \epsilon^6 $\\
\hline
$g_{15}$      & 3.75112   & $-0.120811 \epsilon ^3-0.314547 \epsilon ^4-0.616424 \epsilon ^5-0.985647 \epsilon ^6 $ \\
\hline
$g_{18}$     & 9.27125  & $-0.0215131 \epsilon ^3-0.0332408 \epsilon ^4-0.0501992 \epsilon ^5-0.057848 \epsilon ^6 $ \\
\hline
\end{tabular}
\caption{Numerical computation of GPLs.}
\label{tab:polylog3}
\end{table}

\section{Conclusion}
Through the calculation of pentagon families with up to three massive legs and massless propagators, we have demonstrated the ability of the SDE approach in handling multiscale FI and obtaining analytical results in terms of GPLs. We have presented solutions of the canonical DE for each pentagon family up to weight six, but the closed form of the boundary terms allows one to trivially obtain higher-weight solutions. We have also presented numerical results and timings for the evaluation of our solutions in a Euclidean point. Obtaining fast numerical results through analytical expressions in physical points is an open problem, due to the challenging task of analytically continuing the resulting GPLs when algebraic letters are present in the alphabet. In this contribution we have focused on integrals with massless propagators. Recently a step forward has been made in the study of pentagon integral families involving one internal mass using the SDE approach \cite{Syrrakos:2021nij}.

\section*{Acknowledgements}
NS would like to thank the organisers of RADCOR 2021 for the opportunity to present recent research results.


\paragraph{Funding information}
This research is co-financed by Greece and the European Union (European Social Fund- ESF)
through the Operational Program Human Resources Development, Education and Lifelong
Learning 2014 - 2020 in the context of the project "Higher order corrections in QCD with applications to High Energy experiments at LHC" -MIS 5047812. NS work was supported by the Excellence Cluster ORIGINS funded by the Deutsche Forschungsgemeinschaft (DFG, German Research Foundation) under Germany's Excellence Strategy - EXC-2094 - 390783311.





\nolinenumbers

\end{document}